\documentstyle[12pt]{article}
 \advance\oddsidemargin by -1in
\advance\evensidemargin
 by -1in
\advance\topmargin by -1.5cm
\makeindex

\newcommand\beq{\begin{equation}}
 \newcommand\eeq{\end{equation}}
\newcommand\beqn{\begin{eqnarray}}
 \newcommand\eeqn{\end{eqnarray}}
\newcommand{\doublespace} {
 \renewcommand{\baselinestretch} {1.6}
\large\normalsize}
 \def\sel{\sigma_{el}^{VN}}
 \def\sin{\sigma_{in}^{VN}} \def\stot{\sigma_{tot}^{VN}}
 \def\inf{\int_{-\infty}^{\infty}}

\begin{document}
\vspace*{0.5cm}
\hspace*{9cm}{\Large\bf MPIH-V41-1995}\\
\\
\hspace*{9.5cm}{\Large\bf nucl-th/9511018}
\vspace*{1.5cm}

 \centerline{\huge \bf Where is the Baseline}
\medskip
 \centerline{\huge \bf for Color Transparency Studies}
\smallskip
 \centerline{\huge \bf with Moderate Energy}
\smallskip
 \centerline{\huge {\bf Electron Beams?}\footnote{Based on the  talks
 presented by B.~Kopeliovich at the Workshop
{\it Options for Color Coherence/Transparency
 Studies at CEBAF}, CEBAF, May 22-24, 1995, at
the {\it ELFE} (Electron Lab. For Europe) Workshop, Cambridge, July 22-29,
1995,
and at the Workshop {\it Role of the Nuclear Medium in High-Energy Reactions},
Trento, September 22 - October 6, 1995}}

\vspace{.5cm}
\begin{center}
{\Large Boris~Kopeliovich}\footnote{On leave from Joint Institute for
 Nuclear Research, Laboratory of Nuclear Problems,
\newline Dubna, 141980 Moscow Region, Russia.  E-mail:
 bzk@dxnhd1.mpi-hd.mpg.de}

\vspace{0.3cm}

 {\sl Max-Planck Institut f\"ur Kernphysik, Postfach 103980,
\newline 69029 Heidelberg, Germany}\\

\vspace{0.3cm}

 {\Large Jan Nemchik}\footnote{On leave from Institute of Experimental Physics
 SAV, Solovjevova 47, CS-04353 Kosice, Slovakia}

\vspace{0.3cm}

 {\sl Dipartimento di Fisica Teorica, Universit\`a di Torino\\ I-10125, Torino,
 Italy}
\end{center}

\vspace{0.5cm}
\begin{abstract}

 Study of color transparency (CT) effects at moderate
 energies is more problematic than is usually supposed.
Onset of CT can be imitated  by other mechanisms,
which contain no explicit QCD dynamics. In the case of the $(e,e'p)$ reaction
the standard inelastic shadowing well known in the pre-QCD era, causes a
substantial growth of nuclear transparency with $Q^2$ and a deviation from the
Glauber model, analogous to what is assumed to be a signal of CT.

 In the case of exclusive virtual photoproduction of vector mesons, CT is
expected to manifest itself as an increase of nuclear transparency with $Q^2$
in production of the ground states and as an abnormal nuclear enhancement for
the radial excitations.  We demonstrate that analogous $Q^2$-dependence
can be caused at moderate energies
by the variation of so called coherence length,
which is an interference effect,
even in the framework of the vector dominance model.

One should disentangle the real and the mock CT effects
in experiments planned at CEBAF, at the HERMES spectrometer,
or at the future electron facility ELFE.

\end{abstract}

\newpage

\doublespace
\section{Introduction}
Color transparency phenomenon (CT)
is a manifestation of the color
dynamics of strong interaction which was predicted
to occur in diffractive
interaction with nuclei \cite{zkl,bbgg,knnz}, in quasielastic
high-$p_T$ scattering of electrons and hadrons off nuclei
with high momentum transfer
\cite{mueller,brodsky}, in quasi-free charge-exchange scattering
\cite{kz87,kz91}.
Unfortunately very few experiments were able to claim confirmation of
CT. The most clear signal of CT was observed with high statistics
in the PROZA experiment at Serpukhov \cite{nur} in quasifree
charge-exchange pions scattering off nuclei at $40\ GeV$
(see discussion in \cite{kz87,kz91}). The observation
of the onset of CT in $Q^2$-dependence of nuclear transparency
in virtual diffractive photoproduction of $\rho$-mesons was
claimed recently by the E665 collaboration at Fermilab \cite{e665}.
Unfortunately the  statistical confidence level of the signal
is quite poor.

\medskip

The important signature of CT is the rising nuclear transparency
(vanishing final state interaction) with increasing hardness of
the reaction. However, available experimental
facilities do not still allow to reach
that kinematical region where
a strong signal of CT is expected, but only a onset of of this phenomenon.
In such circumstances one should be cautious about effects which may
mock a signal of CT. In view of the  weakness of the expected
signal of CT one has to understand the origin of the ''background'', i.e.
to know what to expect in absence of the CT effect.

In present paper we would like to draw attention to some effects
which do not rely upon the QCD dynamics, but can mock the onset of CT.
We discuss such effects in
quisielastic electron scattering and in
electroproduction of vector mesons off nuclei.

\section{$Q^2$-dependence of nuclear transparency in $(e,e'p)$}

 Recent failure of the NE18 experiment at SLAC \cite{ne18} to observe the onset
 of color transparency (CT) in $(e,e'p)$ reaction
 has excited interest to the baseline for such a study. It was realized
 that even the Glauber model have a substantial uncertainty.
Nevertheless,  a nearly
$Q^2$-independent nuclear
 transparency is expected in the Glauber approximation,  what makes it
 possible to single out the $Q^2$-dependent effects \cite{mueller,brodsky}.

We call the Glauber eikonal approximation an approach disregarding
any off diagonal diffractive rescatterings of the ejectile, which itself
is assumed to be just a proton.

\medskip

\begin{figure}[tbh]
\includegraphics{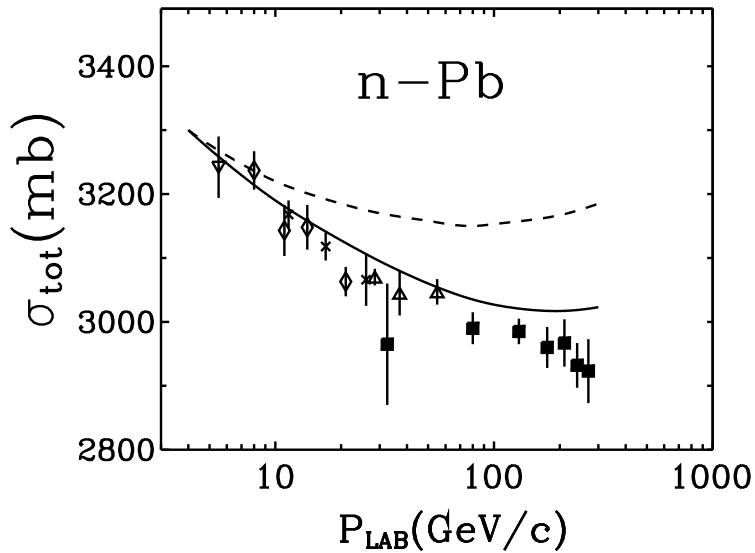}
\begin{center}
\vspace{8cm}
\parbox{13cm}
{\caption[Delta]
{Data on $n-Pb$ total cross
section (ref. \cite{murthy} and references therein).
The dashed curve is corresponds to the Glauber approximation.
The solid line shows the effect of inclusion of the first order
IC to the total cross section
as it is calculated in ref. \cite{murthy}}
\label{fig1}}
\end{center}
\end{figure}

 It is known since Gribov's paper \cite{gribov}, that
the  Glauber model should be corrected for inelastic shadowing at high
energies.
The very existence and the numerical evaluations of the inelastic corrections
(IC)
was nicely confirmed by the high precision measurements of the total
cross sections of interaction of neutrons \cite{murthy} and neutral K-mesons
\cite{gsponer} with nuclei. Due to IC the total cross section
turns out to be smaller, i.e. nuclear matter is more transparent, than is
expected in the Glauber approximation. Important for further discussion is
the fact that the deviation (IC) from the Glauber model grows with energy.
An example is shown in fig. \ref{fig1}. The data for $n-Pb$
total cross section as function of energy are compared
with the
Glauber approximation corrected or not for the
IC, evaluated in \cite{murthy,kk}.

According to these results one can expect that nuclear
matter becomes more transparent at higher ejectile energy, or
$Q^2$ because
the energy $\nu = Q^2/2m_N$ correlates with $Q^2$ within the
quasielastic peak.
Such a rising $Q^2$-dependence of the nuclear transparency
can imitate the CT effects \cite{kn-cebaf}, which are expected to
manifest themselves as a monotonous
growth of the nuclear transparency with
$Q^2$ \cite{mueller,brodsky}.

\medskip

\begin{figure}[tbh]
\includegraphics{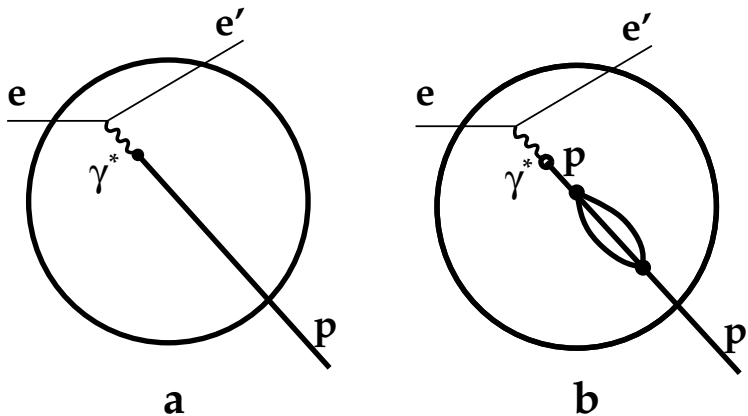}
\begin{center}
\vspace{6cm}
\parbox{13cm}
{\caption[Delta]
{Cartoon showing $A(e,e'p)A'$ reaction with eikonal elastic final state
interactions (a) and with diffractive production of inelastic intermediate
state (b)}
\label{fig2}}
\end{center}
\end{figure}

At this point it worth reminding that CT
is a particular case of Gribov's inelastic
shadowing, provided that QCD dynamics
tunes many elastic and inelastic diffractive
rescatterings in
the final state to cancel each other  \cite{zkl,k-rev,jm} at high $Q^2$.
Therefore, one may
think that there
is no sense in picking up only one IC
from many others, which all together
build up CT. However,
searching for CT effects one should ask himself first of all,
what happens
if CT phenomenon does not exist; for instance, if the ejectile
in $(e,e'p)$ reaction on a bound nucleon is not a small-size wave
packet, but is a normal
proton. There is a wide spread opinion that the
correct answer is provided by the Glauber model.
However, the IC shown schematically in fig. \ref{fig2},
makes the nuclear
matter more transparent.

This first order IC corresponds to the diffractive
production of inelastic intermediate states by the ejectile proton
while it propagates through the nucleus.
  The proton waves with and without this
 correction interfere with each other, while the contributions from
different production points add up incoherently
because the momentum transfer
in the $(e,e'p)$ reaction is large.
It is important that IC has a positive relative sign, provided that all
diffraction amplitudes are imaginary \cite{kl,jk,fgms}.
The resulting nuclear transparency reads,

 \beqn
Tr(Q^2) & = & \int d^2b \int_{-\infty}^{\infty}
 dz\rho(b,z)\ \exp[-\sigma^{NN}_{in} \int_z^{\infty} dz'
\rho(b,z')]\times\nonumber\\
& & \left[1+4\pi \int dM^2 \frac{d\sigma}{dM^2dt}|_{t=0}\
 F^2_A(b,z,q_L)\right]^2
\label{2}
\eeqn

 Here $b$ and $z$ are the impact parameter and the longitudinal coordinate of
the
 bound proton which absorbs the virtual photon.  $\sigma^{NN}_{in}$ is
 inelastic $NN$ cross section. We assume that the $(e,e'p)$
cross section is integrated
over the transverse momentum of the ejectile proton relative
to the photon direction, and over the missing momentum, which is the difference
between the photon and the proton momenta.
$T(b)=\int_{-\infty}^{\infty}dz \rho(b,z)$ is
 the nuclear thickness function.  $d\sigma/dM^2dt|_{t=0}$ is the forward
diffraction dissociation cross
 section in $NN$ interaction.  $F_A(b,z,q_L)$
 is the nuclear longitudinal form factor \cite{kk},

 \beq
F_A(b,z,q_L)=\int_{z}^{\infty}dz'\ \rho(b,z')e^{iz'q_L}\ ,
\label{3}
\eeq
where $q_L=(M^2-m_N^2)/2\nu$ is the longitudinal momentum transfer in the
diffraction dissociation. This form factor is of special importance,
because it provides the $Q^2$-dependence of
 nuclear transparency.

\begin{figure}[tbh]
\includegraphics{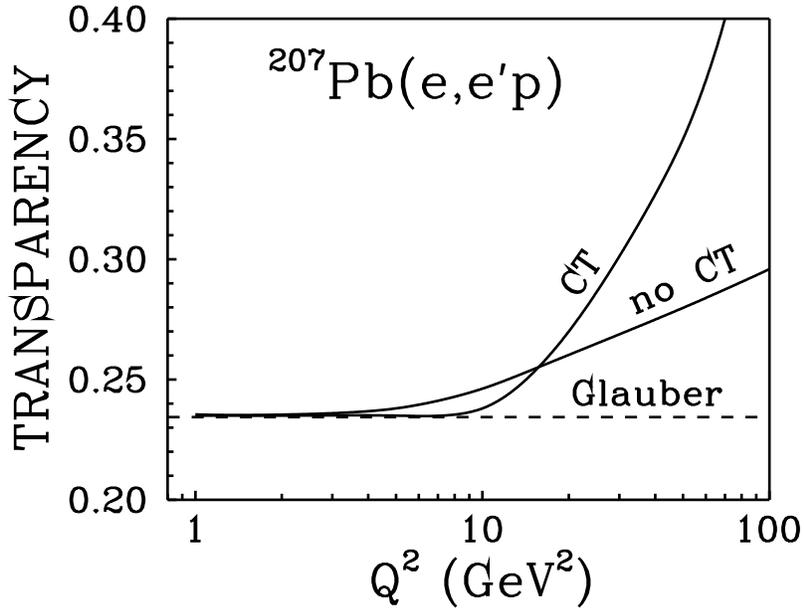}
\begin{center}
\vspace{9cm}
\parbox{13cm}
{\caption[Delta]
{Comparison of the Glauber model (dashed line) with the
model \cite{nnz} incorporating with CT (solid curve -- CT)
and with our calculation of the first-order IC using eq.(\ref{2})
(solid curve
-- no CT)}
\label{fig3}}
\end{center}
\end{figure}

The detailed calculation of an analogous IC
to the nuclear total cross section was performed in \cite{murthy}.
We use the same parameterization of the data on
$d\sigma/dM^2dt$ as in \cite{murthy} and the realistic nuclear density from
\cite{jager} to calculate expression (\ref{2}).  Following refs.
\cite{kk,murthy,gsponer} we assume that
the inelastic intermediate states attenuate in nuclear medium
with the same inelastic cross section as the proton. The predicted
growth of nuclear transparency with $Q^2$ in $Pb(e,e'p)$
is compared in fig. \ref{fig3}
with what is expected to
be the onset of CT \cite{nnz}. We use $\sigma^{NN}_{in} =
33\ mb$ in order to have the same transparency in the Glauber approximation
as in ref. \cite{nnz}. We see that
these two mechanisms, one with and another one without CT dynamics
predict about the same magnitude of deviation from the Glauber model
up to about $Q^2\approx 20\ GeV^2$. It is especially difficult to disentangle
the real CT effects and the first-order IC
because of a substantial model-dependence
of the theoretical predictions for CT. In order to
detect reliably a signal of CT one needs $Q^2$ at least of a
few tens of $GeV^2$, where
 the growth of transparency provided by the first
IC saturates at quite a low level.

\begin{figure}[tbh]
\includegraphics{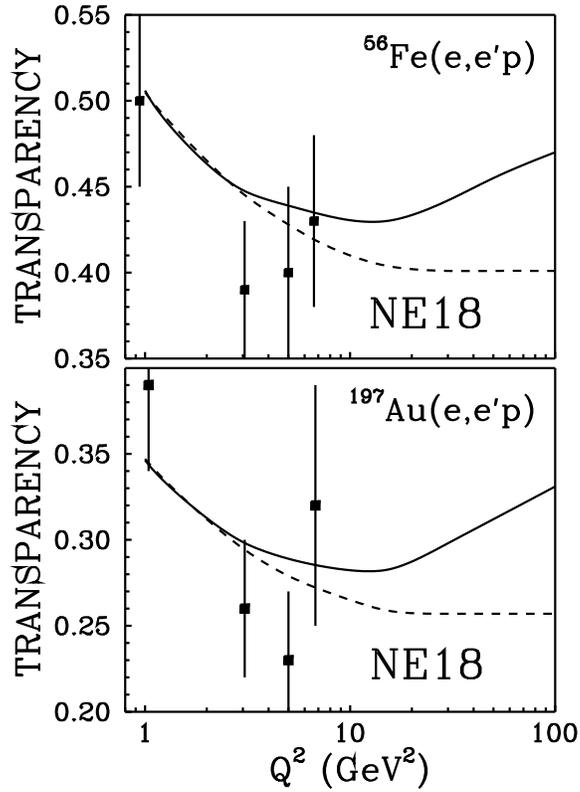}
\begin{center}
\vspace{11cm}
\parbox{13cm}
{\caption[Delta]
{Comparison of the Glauber model (dashed line) and of the results
of our calculations of the standard first-order
IC, eq. (\ref{2}), (solid line)
with the data from the NE18 experiment \cite{ne18}}
\label{fig4}}
\end{center}
\end{figure}

Our calculations are compared with the data from the
NE18 experiment at SLAC \cite{ne18} in fig. \ref{fig4}.
We use a realistic $\sigma_{in}^{NN}$ from ref. \cite{compil} which exhibit
a decreasing energy-dependence at low energies (compare with \cite{fsz}).
Of course, more sophisticated calculations may consider the effects of Fermi
motion \cite{jk,bbk,kolya2}, few-nucleon correlations
\cite{benhar1,kolya2,kohama,rinat,moniz},
accuracy of the closure approximation \cite{rj}, etc. We try to escape these
complications to make the presentation simpler and clearer.
The relative contribution of the IC is expected to be nearly
independent of the mentioned above details of nuclear structure.

Note that we predict a bigger effect of inelastic shadowing than that in
the total hadron-nucleus cross sections \cite{murthy,gsponer}.
In the latter case case it is the correction to the small exponential term
in the elastic amplitude which is subtracted from unity,
while in the present case we deal with a net
transparency effect.

 \medskip

To conclude this section, we estimated the first-order IC, which
causes a growth of nuclear transparency with $Q^2$
in quasielastic scattering of electrons
off nuclei and can imitate
the onset of CT up to $Q^2\sim 20\ GeV^2$.
 The evaluation of this IC
is independent of our ideas about
QCD dynamics of hard interaction, since it is based only on the data on
diffractive dissociation. Although this correction is a part of the
total CT pattern, it survives any modification of the underlying dynamics and
should be considered as a {\it baseline} for CT studies.
 One can reliably disentangle this
contribution and the real CT effect only at $Q^2$ of a few tens of $GeV^2$,
where
the former saturates, but CT provides a growth of nuclear
transparency up to unity.
Note that in order to suppress the IC under discussion, one can use
light nuclei, at the expense of a smaller CT effect.

A new possibility to search for CT effect, which is free of
the contribution of the IC discussed above,
was suggested in \cite{jk,bbk}.
The asymmetry of nuclear transparency relative to
the longitudinal missing momentum turns out to be sensitive
to the QCD dynamics of quasi-elastic electron scattering on a bound
nucleon and reflects generic properties of CT. A more realistic evaluation of
this effect
\cite{bnz-ct} demonstrates that this is a promising way to detect a CT signal
in a high-statistics experiment.

\section{Exclusive electroproduction of
vector mesons}

Electrons are a source of energetic virtual photons. In the
reaction of virtual diffractive photoproduction of
vector mesons on can control the size of the produced
 wave packet varying $Q^2$, but keeping the photon energy fixed.
Therefore, the inelastic corrections considered in the previous section
are irrelevant, since they are only energy-dependent.

Although the collaboration E665 \cite{e665} has observed a signal of CT
 effects in the exclusive muoproduction
of $\rho$-mesons predicted in \cite{knnz},
one should be cautious
 with the conclusions. The predictions for $Q^2$-dependence of
nuclear transparency \cite{knnz} were done for
the asymptotically high photon energies.
We demonstrate below that variation of
the coherence length may imitate to some extend the CT effects.

The vector mesons photoproduced at different longitudinal coordinates have
 relative phase shifts $\Delta z q_L$ due to the difference in
 the photon and the meson longitudinal momenta $q_L=(Q^2+m_V^2)/2\nu$. For this
 reason only those mesons interfere constructively which are produced
 sufficiently close to each other: $\Delta z\leq l_c$, where

 \beq l_c={1\over q_L}=\frac{2\nu}{Q^2+m_V^2}
\label{4}
\eeq is called coherence length.

One can provide  this with a space-time interpretation. The photon develops a
hadronic fluctuation which can interact with the nucleus during its lifetime
$t_c=l_c/c$ ($c$ is the speed of light). If $l_c$ is much shorter than the mean
internucleon distance, there is obviously no nuclear
shadowing in the initial state. On the other hand, if $l_c$ is much longer than
the mean free path of the vector meson in the nuclear medium or the nuclear
radius,
one cannot any more distinguish between the initial and the final state
interactions.
Nuclear shadowing is expected in this limit
to be the same as in the meson-nucleus interaction. Thus, the energy- and
$Q^2$-variation of $l_c$ may result in substantial changes in the nuclear
shadowing,
which can be easily mixed up with the CT effects in some cases
\cite{hkn-cebaf}.  For the light vector meson
($\rho, \omega,\phi$) production the transition region covers the
energies from a few to few tens GeV. For charmonium production the
corresponding
energy range is an order of magnitude higher.

We demonstrate below how this space-time pattern is realized in the framework
of the formal Glauber theory, which does not utilize any
space-time development.

\subsection{Coherent electroproduction of the ground states off nuclei.
Glauber model}

We call coherent production the process where the target
nucleus remains intact, so all the vector mesons produced
at different longitudinal coordinates and impact
parameters must add up coherently. This condition substantially simplifies
the expression for the photoproduction cross section,
which reads \cite{bauer},

 \beq
Tr^{coh}(\gamma^*A\rightarrow VA)=\frac{(\stot)^2}{4\sel} \int d^2b\left |\inf
dz\ \rho(b,z)\ e^{iq_Lz} \exp\left[-{1\over 2}\stot\int_z^{\infty}
dz'\rho(b,z')\right ]\right |^2
\label{5}
\eeq
We use hereafter the optical approximation for the sake of simplicity,
which is quite precise for heavy nuclei.

 For numerical calculations we use $\sigma_{tot}^{\rho N} = 25\ mb$
and $\sigma_{tot}^{\phi N} = 17\ mb$.
The elastic cross section is estimated by means of relation,
 $\sel\approx (\stot)^2/16\pi B^{VN}$, where the slope parameter was fixed at
 $B^{VN}=8\ GeV^{-2}$ and $7\ GeV^{-2}$ for $\rho$ and $\phi$ respectively.

\begin{figure}[tbh]
\includegraphics{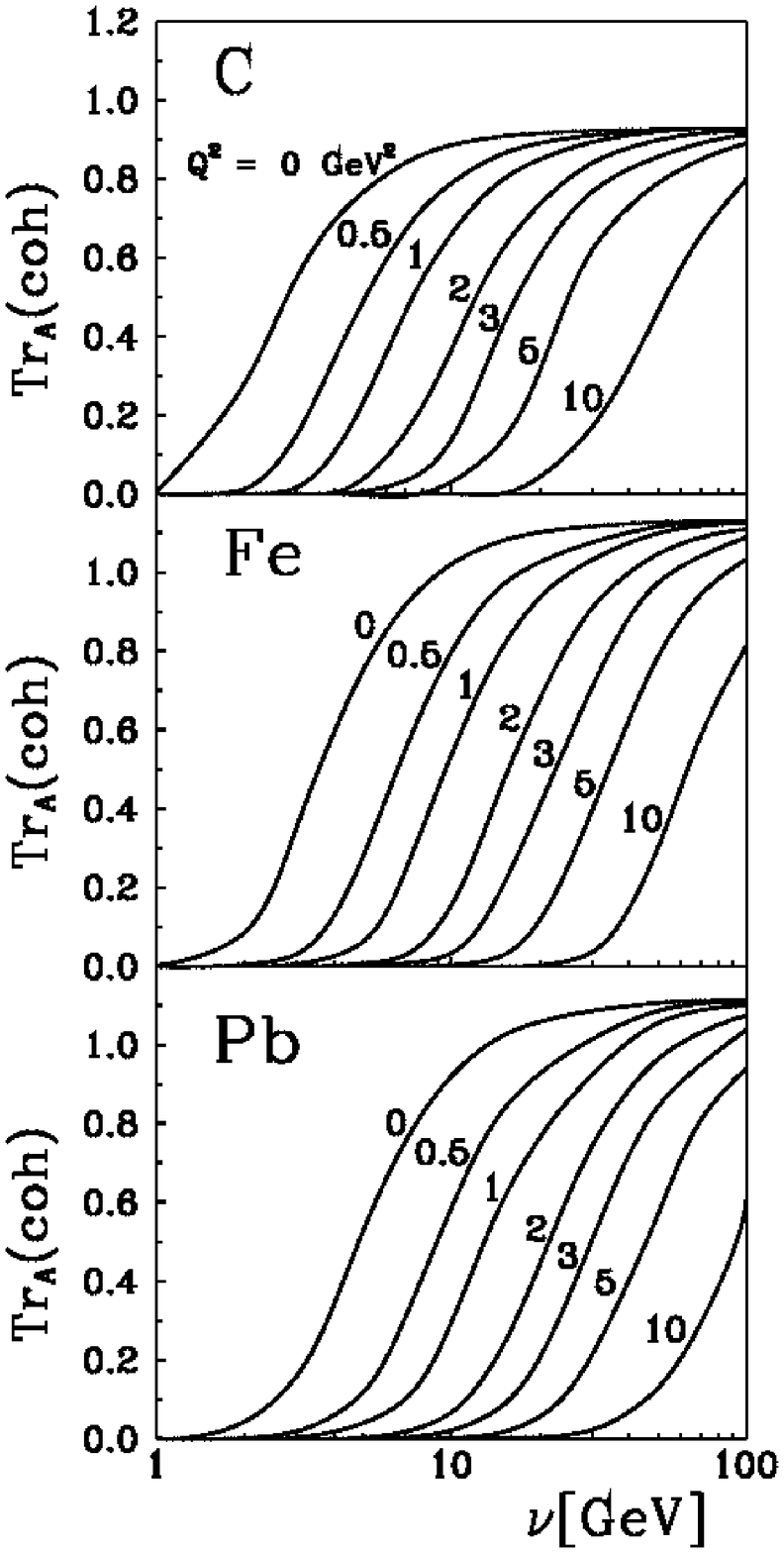}
\includegraphics{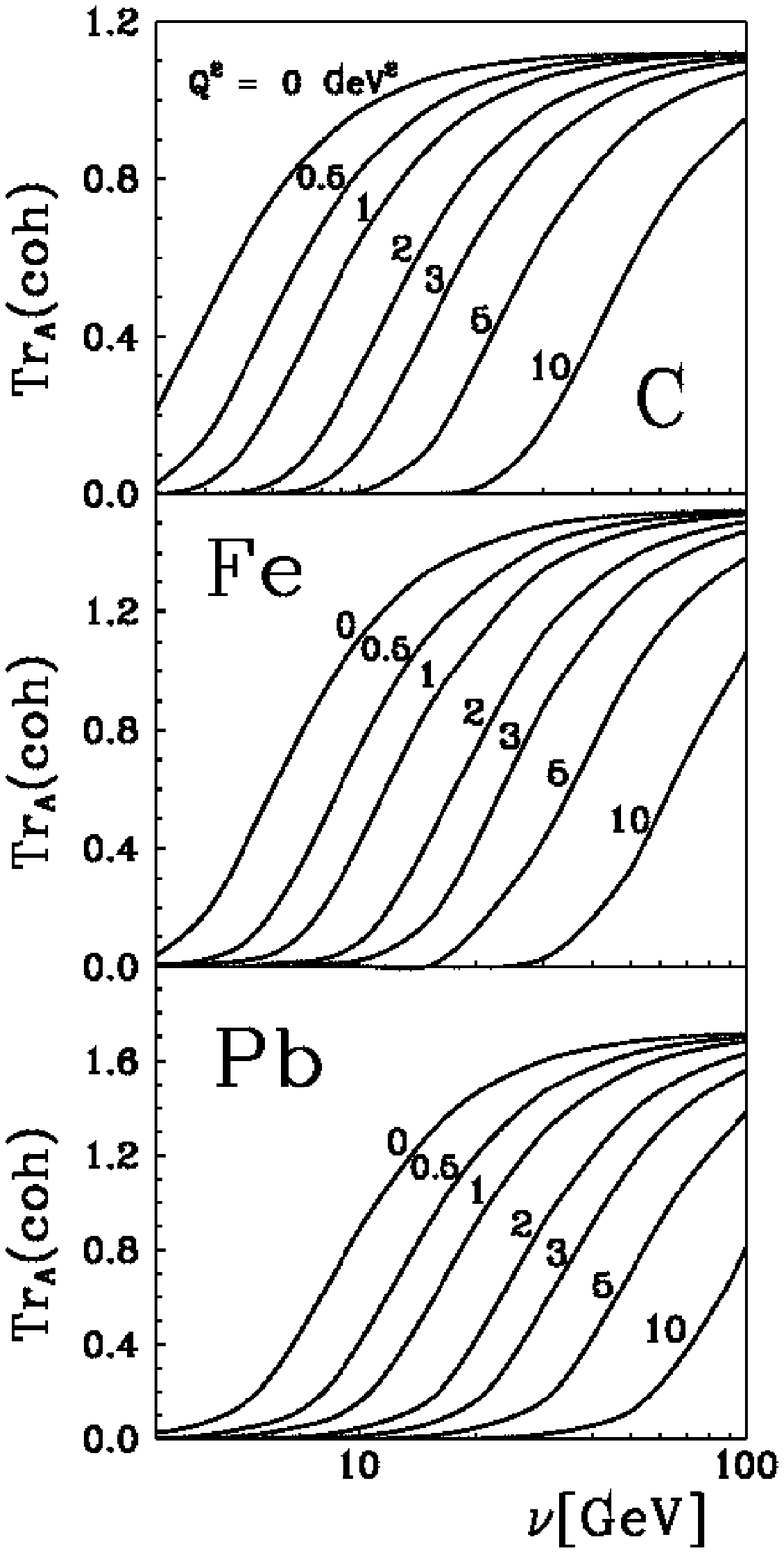}
\begin{center}
\vspace{15cm}
\parbox{13cm}
{\caption[Delta]
{Energy dependence of nuclear transparency in coherent photoproduction
of $\rho$-mesons on nuclei versus the photon virtuality, $Q^2$. The curves are
the results of Glauber model calculations using eq. (\ref{5})}
\label{fig5}}
{\caption[Delta]
{The same as in fig. \ref{fig5} but for production of $\phi$-mesons}
\label{fig6}}
\end{center}
\end{figure}

 The results of calculation of the energy dependence of the
nuclear transparency for coherent photoproduction of $\rho$ and $\phi$ mesons
integrated over momentum transfer are plotted in figs.
\ref{fig5} and \ref{fig6} for
different photon virtualities.
We see
that even the $\rho$-meson coherent production on medium and heavy nuclei
is strongly suppressed at CEBAF
energies, but is quite a sizeable effect at energies of HERMES-ELFE.

\begin{figure}[tbh]
\includegraphics{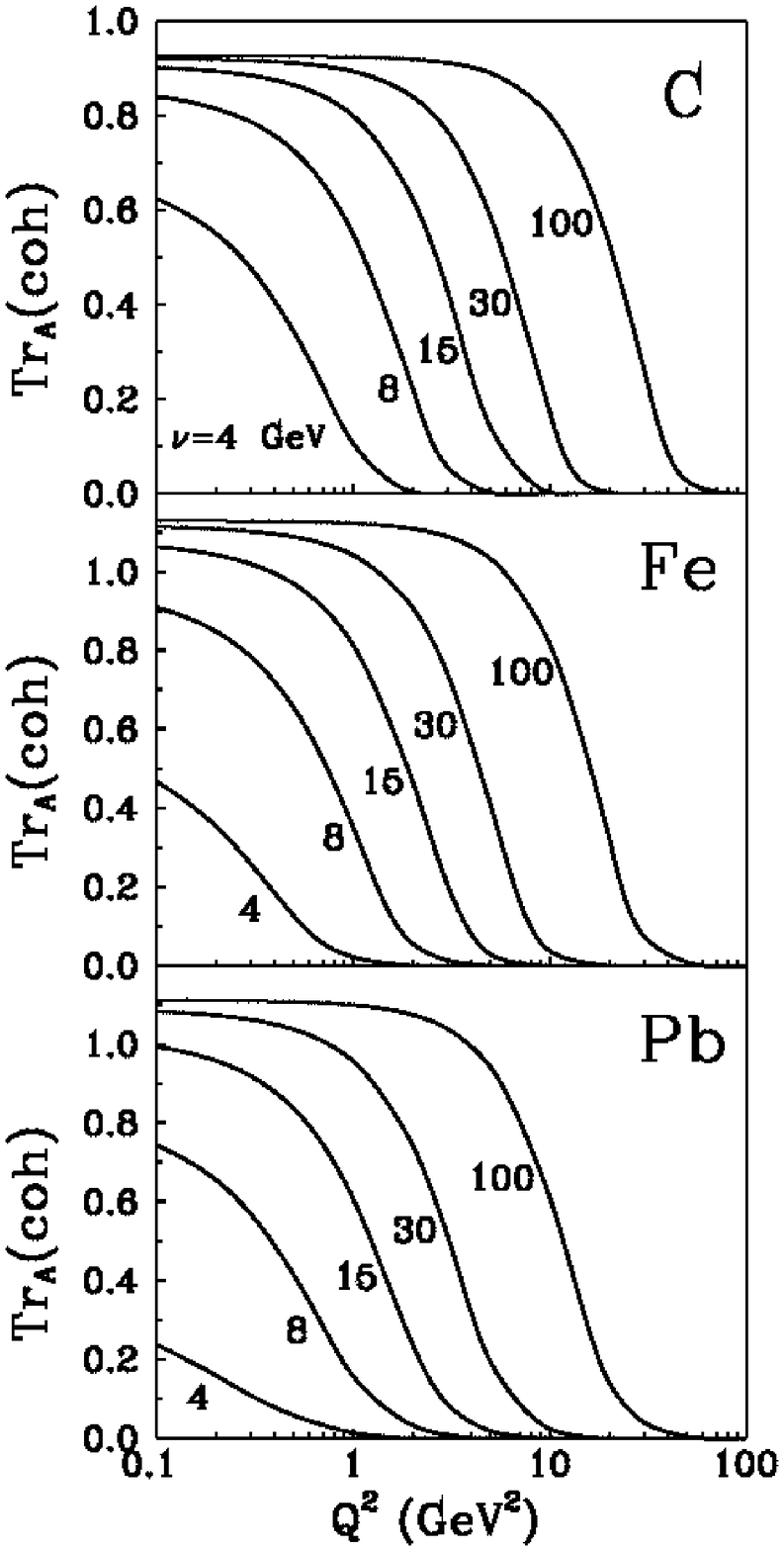}
\begin{center}
\vspace{15cm}
\parbox{13cm}
{\caption[Delta]
{$Q^2$-dependence of nuclear transparency of coherent photoproduction
of $\rho$-meson versus photon energy, calculated with eqs.
(\ref{6})--(\ref{8}). The figures at the curves show
values of $\nu$ in $GeV$}
\label{fig7}}
\end{center}
\end{figure}

Since the $Q^2$-dependence of the nuclear transparency is usually used
as a signature for CT, we present in fig. \ref{fig7}
the Glauber model predictions for
$Q^2$-variation of the nuclear transparency in the
$\rho$-meson photoproduction at different photon energies.

\begin{figure}[th]
\includegraphics{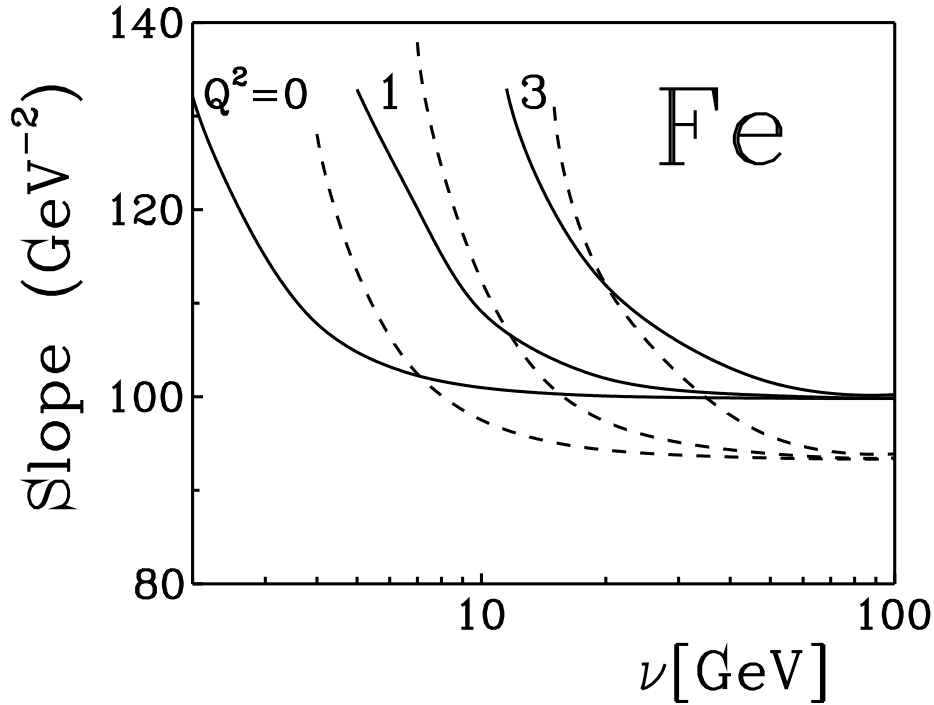}
\begin{center}
\vspace{9.5cm}
\parbox{13cm}
{\caption[Delta]
{Energy dependence of the slope-parameters for coherent photoproduction
of $\rho$ (solid lines) and $\phi$ (dashed lines) vs the photon virtuality,
$Q^2=0,\ 1,\ ,3\ GeV^2$, shown by figures at the curves.}
\label{fig8}}
\end{center}
\end{figure}

The energy variation of the slopes of differential cross section of coherent
 photoproduction of $\rho$ and $\phi$ mesons on iron,
$B_{VA}=1/2\langle b^2\rangle$, is presented in fig. \ref{fig8}.
The condition of coherence results in large values of
$B_{VA} \approx R_A^2/3$.

\medskip

The energy-dependence of the slopes demonstrates a crossing behaviour:
 $B_{\rho A}< B_{\phi A}$ at low energy,
but  $B_{\rho A}> B_{\phi A}$ at high energies.
This can be understood as follows. At high
energy the photoproduction cross section is proportional to the elastic
one for $V-A$ interaction.
In this regime a heavy nucleus is almost black
at small impact parameters. At the same time,
the partial elastic amplitude at the nuclear periphery is proportional
to the $V-N$ elastic amplitude, i.e. it is smaller
for $\phi$ than for $\rho$. This results in a smaller slope parameter
for $\phi$ than for $\rho$.

At low energies situation changes, everything
is produced on the back surface of the nucleus.
This leads to a rising relative
contribution of the nuclear periphery, i.e. to a growth of the slope parameter.
Since the mean free path of the $\rho$ is shorter, the contribution of the
small
impact parameter region is more suppressed, than for the $\phi$-meson.
Therefore, the $\rho$ slope parameter is larger than that of the $\phi$-meson
in this energy limit.

\subsection{Incoherent electroproduction of the ground states off nuclei}

The incoherent diffractive production is associated with a break up
of the nucleus, but without production of new particles.

 The correct formula for exclusive incoherent electroproduction of vector
mesons
 incorporating the effects of coherence length is derived for the first
time\footnote{The formula for incoherent photoproduction of vector mesons
presented in ref. \cite{bauer} differs from our eqs. (\ref{6})--(\ref{8}),
and we consider it as incorrect. That formula underestimate available data
on real photoproduction of $\rho$ off nuclei (see corresponding
discussion in ref. \cite{bauer}). Our calculations nicely agree with the data.}
 in ref. \cite{hkn}. The nuclear transparency for the cross section integrated
 over momentum transfer can be represented as a sum of three terms,

 \beq
Tr_{inc}(\gamma^*A\rightarrow VX)=Tr_1+Tr_2-Tr_{coh}\ ,
\label{6}
\eeq
where the first term

 \beq
Tr_1=\frac{1}{\sin} \int d^2b \left [1-e^{-\sin T(b)}\right ]
\label{7}
\eeq
corresponds to the case when the vector
 meson is produced in both interfering amplitudes
on the same nucleon. If those nucleons are different, the
 corresponding term $Tr_2$ in eq. (\ref{6}) reads,

 \beqn
& & Tr_2 = \frac{\stot}{2\sel}(\sin-\sel) \int d^2b\ \inf dz_2\
8/GeV \rho(b,z_2) \int_{-\infty}^{z_2} dz_1\ \rho(b,z_1)\ \times \nonumber\\ &
&
 e^{iq_L(z_2-z_1)}\ \exp\left[ -{1\over 2}\stot\ \int_{z_1}^{z_2}dz
 \rho(b,z)\right] \exp\left[ -\sin\ \int_{z_2}^{\infty} dz\ \rho(b,z)\right]
\label{8}
\eeqn

 Two first terms in eq. (\ref{6}) correspond to the sum over all final
 states of the nucleus, including the case when the nucleus remains in
the ground state.  For this reason the
 third term is subtracted in eq.
 (\ref{6}).

 In the low- and high-energy limits eqs.  (\ref{6})-(\ref{8}) look much
simpler.
At low energies $q_L$ is large, what
causes strong oscillations and cancellations in
eqs. (\ref{8}), (\ref{5}). Only the first term in eq. (\ref{6}) survives, and
this can be interpreted as a result of the shortness of the coherence length:
the
photon penetrates without attenuation deep inside the nucleus and
instantaneously produces the vector meson on some of the bound
nucleons. Absorption of the meson travelling through the nucleus leads to
a suppression of the nuclear transparency.

 At high-energies $q_L\approx 0$ and eq. (\ref{6}) takes the form,

\beq
Tr_{inc}^{q_L\rightarrow 0}(\gamma^*A\rightarrow VX) \rightarrow
\frac{1}{\sel} \int d^2b
 \left[ e^{-\sin T(b)}-e^{-\stot T(b)} \right]\ =
 \frac{\sigma^{VA}_{Qel}}{\sel}
\label{9}
\eeq

We conclude that the nuclear effects in the high-energy photoproduction of
 vector meson are the same as in the
quasielastic scattering of this meson off
 the nucleus. This result has a clear space-time
interpretation. At high energies the coherence length is long and the
 photon converts into the vector meson long in
advance of the interaction. This is
 usually interpreted in terms of vector dominance model (VDM),
however we {\it did not use}
any assumption of VDM.

\begin{figure}[tbh]
\includegraphics{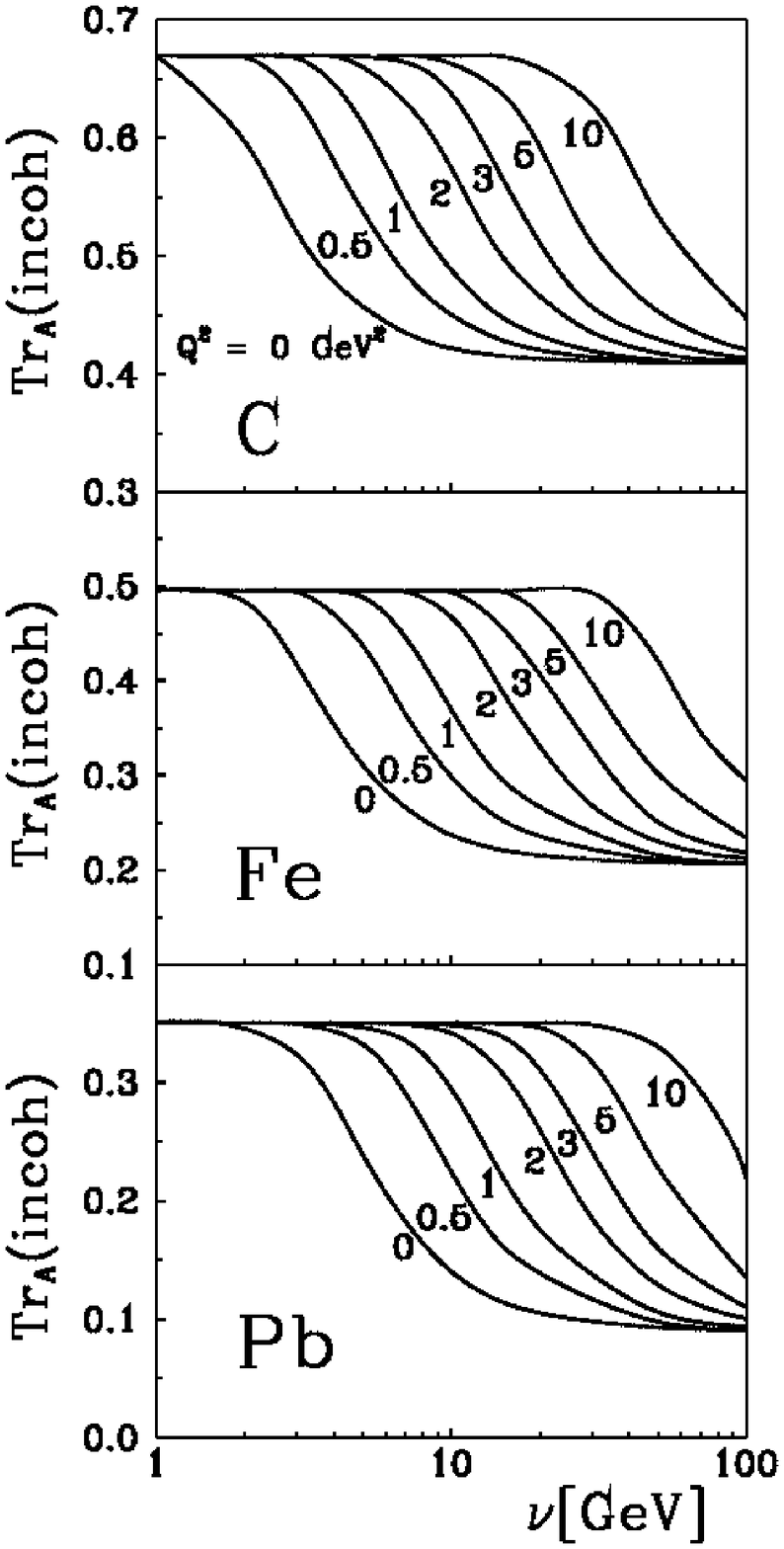}
\includegraphics{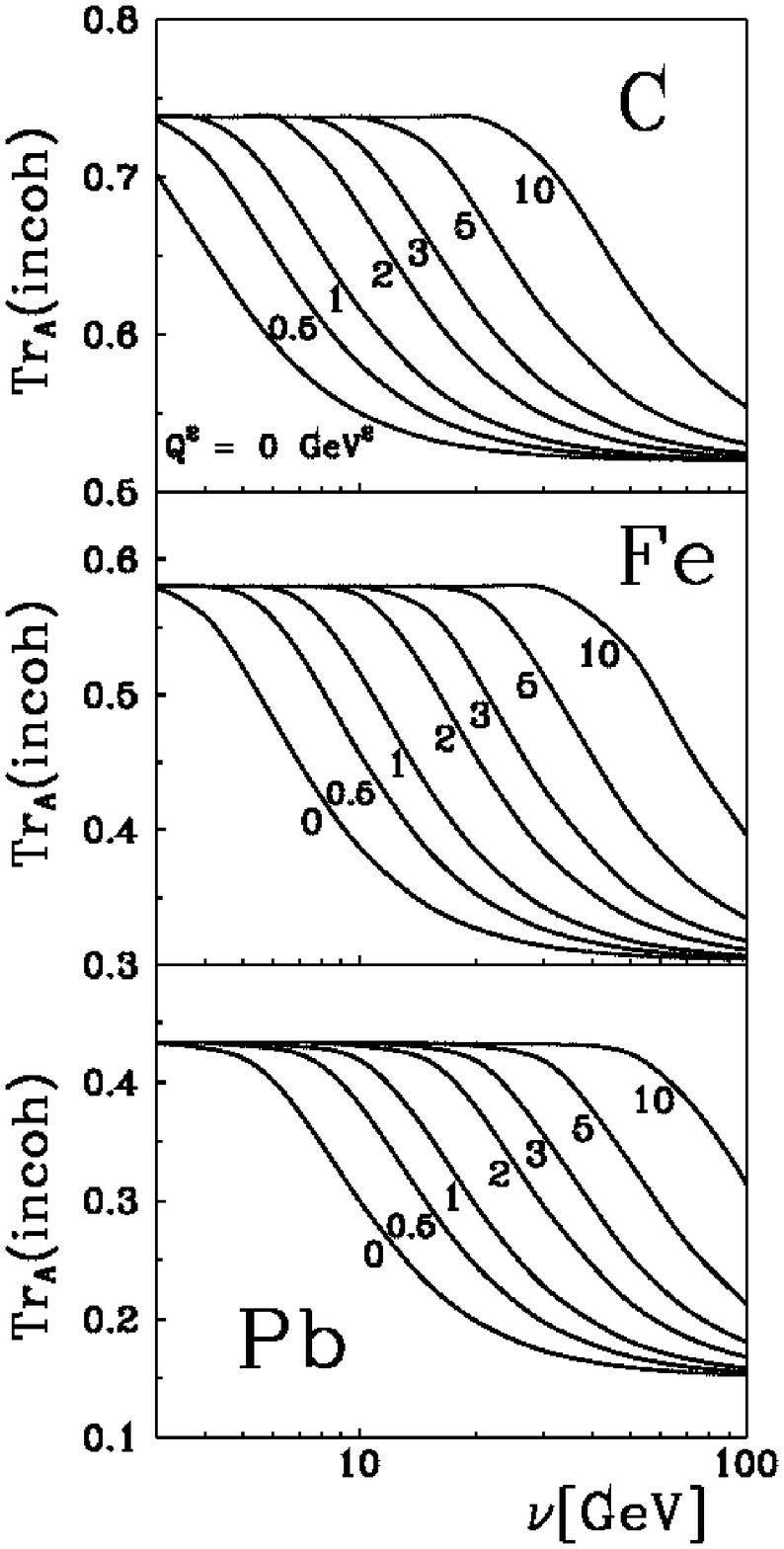}
\begin{center}
\vspace{15cm}
\parbox{13cm}
{\caption[Delta]
{Energy dependence of nuclear transparency for incoherent photoproduction
of $\rho$-mesons on nuclei at different values of $Q^2$,
calculated with eqs. (\ref{6})-(\ref{8})}
\label{fig9}}
{\caption[Delta]
{The same as in fig. \ref{fig9} but for production of $\phi$-mesons}
\label{fig10}}
\end{center}
\end{figure}

The results of calculation of nuclear transparency for
the $\rho$ and $\phi$
incoherent photoproduction are plotted in figs.
\ref{fig9} and \ref{fig10} respectively
as a function of the photon energy and virtuality.

We see that eqs. (\ref{6})-(\ref{8})
predict a dramatic decrease of nuclear transparency from low to high
energies. The explanation follows from the above space-time interpretation.
Namely, the mean length of the path of the meson in the nuclear medium
at high energies is about twice as long as at low energies.
Therefore, one may expect the nuclear transparency at high energies
to be a square of that at low energy. Our results depicted in
figs. \ref{fig9}, \ref{fig10} show that such a simple rule works surprisingly
well.

\begin{figure}[tbh]
\includegraphics{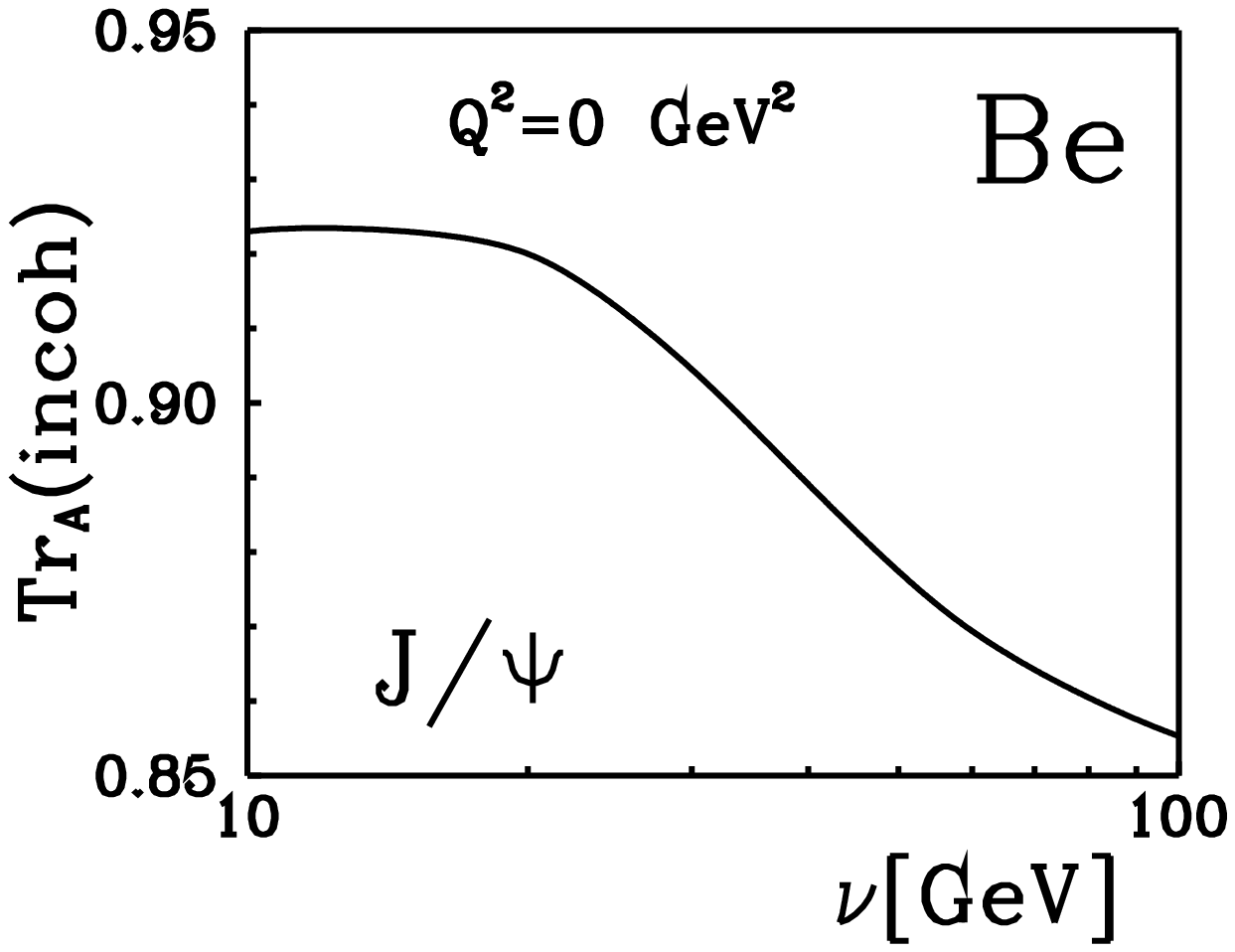}
\begin{center}
\vspace{7cm}
\parbox{13cm}
{\caption[Delta]
{Energy-dependence of nuclear transparency
in the incoherent real photoproduction of $J/\Psi$ on
beryllium, as calculated with eq. (\ref{6})-(\ref{8})}
\label{psi-be}}
\end{center}
\end{figure}

Effects of coherence length for electroproduction of
heavy flavoured vector mesons
are boosted to a higher energy range. However, they are
still important at medium energies for light nuclei.
As an example, we show the energy dependence of the incoherent
real photoproduction of $J/\Psi$ off beryllium in fig. \ref{psi-be}.

\begin{figure}[tbh]
\includegraphics{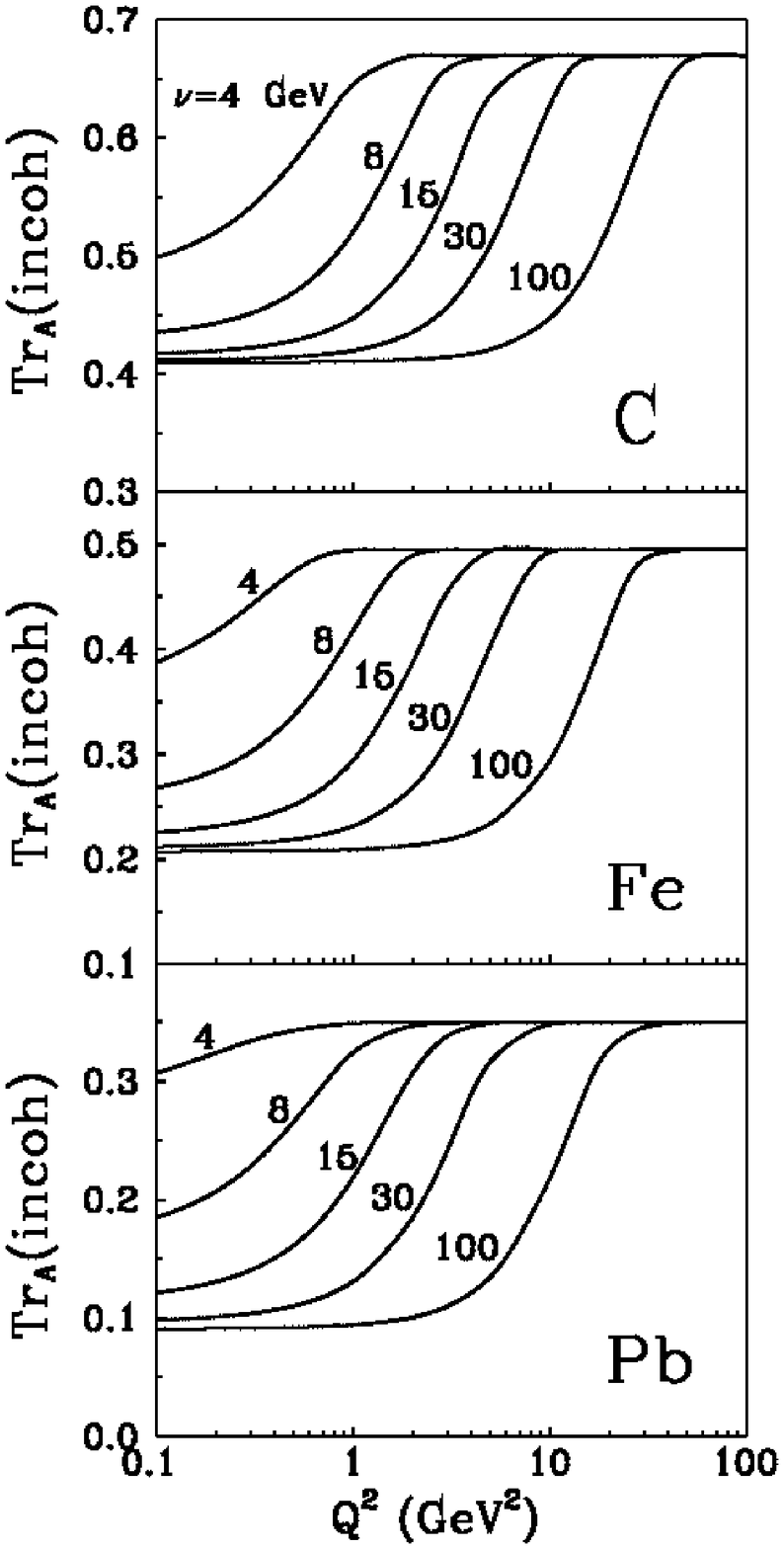}
\begin{center}
\vspace{18cm}
\parbox{13cm}
{\caption[Delta]
{$Q^2$-dependence of nuclear transparency for incoherent photoproduction
of $\rho$-meson at different photon energies $\nu$}
\label{fig11}}
\end{center}
\end{figure}

 Since the coherence length eq. (\ref{4}) is also a function of $Q^2$ at
 fixed energy, the contraction of the coherence length
 causes a growth of nuclear transparency with $Q^2$, which
should be taken into account if one searches for CT effects.  Examples of
 $Q^2$-dependence for $\rho$-meson photoproduction are shown in fig.
\ref{fig11}.
 Nuclear transparency steeply increases and then saturates at high $Q^2$. It is
 not easy to disentangle such a growth of nuclear transparency induced by
the shrinkage of the coherence length and the CT effects. Naively, one could
 hope to search for CT effects at higher $Q^2$, where the Glauber model
predicts
a  saturation of $Tr(Q^2)$. However, the
nuclear transparency cannot reach unity at this energy,
 even in the presence of the CT effects.  $Tr(Q^2)$ saturates at about the same
level as is
 shown in fig. \ref{fig11} both with and without CT effects. Indeed, the
predicted saturation
 signals that the coherence length becomes negligibly short. In this situation
 there is no room for the full CT effect, which needs a destructive
interference of
all the intermediate states up to the masses of the order of $Q$.
In a simplified way this can be interpreted
 as a result of the fast expansion of the produced
small-size $\sim 1/Q$ wave packet up to a certain
size, which depends on the photon energy, rather than
on $Q^2$.

\begin{figure}[tbh]
\includegraphics{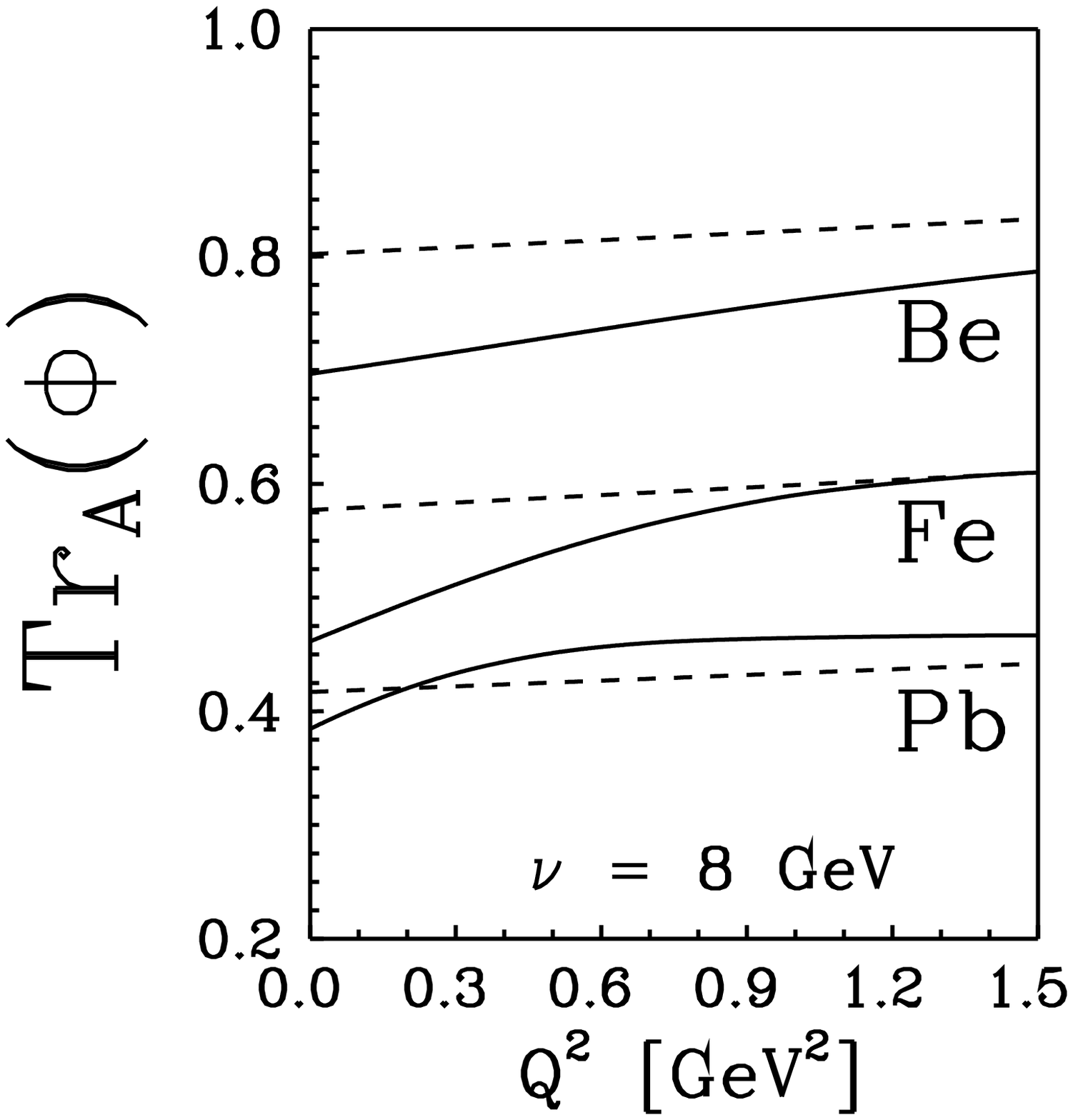}
\begin{center}
\vspace{13cm}
\parbox{13cm}
{\caption[Delta]
{$Q^2$-dependence of nuclear transparency for incoherent photoproduction
of $\phi$-meson at the photon energies $\nu = 8\ GeV$.
The dashed curves are the results of calculations \cite{benhar2}
neglecting the coherence length. The solid curves show our predictions.}
\label{fig12}}
\end{center}
\end{figure}

The importance of the variation of the coherence length is demonstrated
in fig. \ref{fig12} for the $\phi$ photoproduction at $\nu = 8\ GeV$
in comparison with the results of ref. \cite{benhar2} which
neglected the coherence length. Despite the smallness of the
$\phi$ absorption cross section, the coherence length effects are substantial.
Some difference between our and ref. \cite{benhar2} predictions at
high $Q^2$ is caused by the difference in the parameters
for the nuclear density (we use the parameters from \cite{jager}), what,
however, does not affect the $Q^2$-dependence.

\begin{figure}[tbh]
\includegraphics{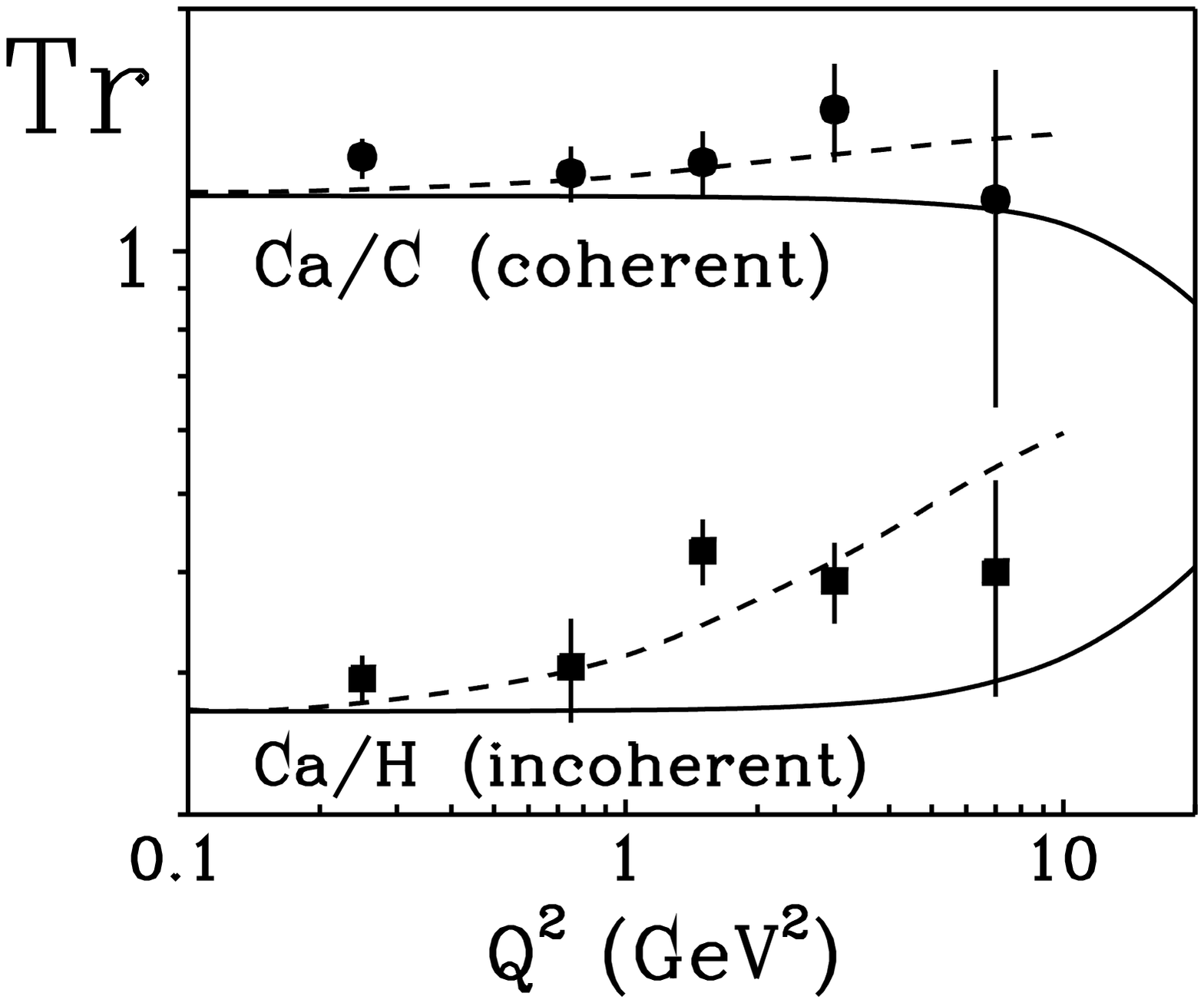}
\begin{center}
\vspace{9.5cm}
\parbox{13cm}
{\caption[Delta]
{$Q^2$-dependence of nuclear transparency for
coherent (upper points and curves) and
incoherent virtual photoproduction of $\rho$-meson.
The solid curves correspond to the Glauber model incorporated
coherence length effects. The dashed curves show CT effects
predicted for asymptotic energies in \cite{knnz}.
The data points are from the E665 experiment \cite{solgado}}
\label{e665}}
\end{center}
\end{figure}

 It is interesting that even the data from the E665 experiment \cite{e665}
 corresponding to mean energy $\langle \nu \rangle = 138\ GeV$ are affected by
 the variation of the coherence length. Our Glauber model predictions for
incoherent
photoproduction of the $\rho$ is
 shown by the bottom solid curve  in fig.  \ref{e665}
in comparison with the E665 data.
We predict a dramatic rise of
 $Tr(Q^2)$ at high $Q^2$. However, the small-$Q^2$ data seem to grow faster
 than our curve,  and they are consistent with the
predicted \cite{knnz} effect of CT.

 In these circumstances the coherent production of vector mesons is maybe a
better
way of searching for CT, since the Glauber model predicts a decreasing
$Q^2$-dependence of the nuclear transparency as is shown by the upper solid
curve in fig. \ref{e665}. Observation of a rising $Tr(Q^2)$
would be a convincing signal of CT. The data are consistent with the
dashed curve \cite{knnz}, which incorporates the CT effects.

\section{Electroproduction of the radial
excitations}

\begin{figure}[tbh]
\includegraphics{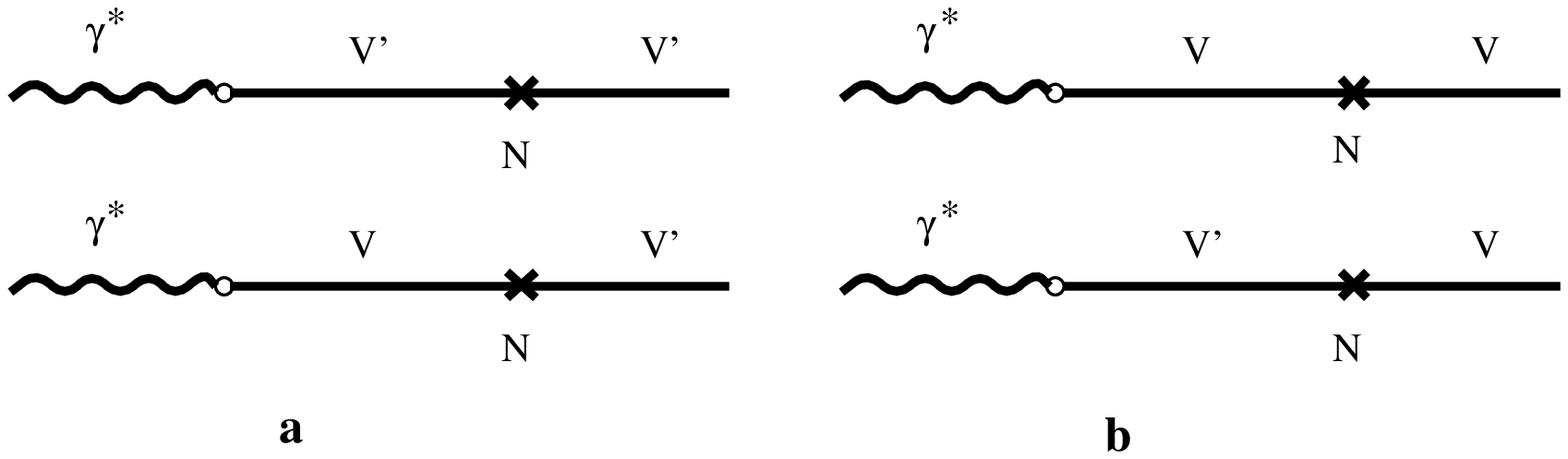}
\begin{center}
\vspace{5cm}
\parbox{13cm}
{\caption[Delta]
{Diagrams for the virtual exclusive photoproduction
of the vector mesons, {\bf a} - the radial excitation $V'$
and {\bf b} - the ground state $V$, in the two channel
approximation}
\label{vv}}
\end{center}
\end{figure}

 As far as it concerns the photoproduction
of the radial excitations $V'$, one cannot
anymore interpret it on the basis of the Glauber eikonal
approximation. Indeed, there are two graphs depicted in fig. \ref{vv}a, which
contribute to the reaction $\gamma^*p\rightarrow V'p$. The second one,
containing the off
diagonal diffractive amplitude $Vp\rightarrow V'p$, cannot be neglected
because the photon coupling to $V$ is much larger than to $V'$.
Therefore, the off-diagonal amplitudes $V \rightleftharpoons V'$
should be added to
the nuclear multiple scattering series as well,
what partially accounts for inelastic
corrections to the Glauber approximation.

 Let us denote the imaginary parts of the
diffractive diagonal and off diagonal amplitudes on a nucleon
$f_{VV},\ f_{V'V'}$ and $f_{VV'}$. We fix the
normalization through the optical
theorem, $f_{VV}=\stot/2$. We use also the notations from ref.  \cite{hk},
$r_V=f_{V'V'}/f_{VV}$ and $\epsilon_V = f_{VV'}/f_{VV}$. Then the ratio of
the photoproduction amplitude of the radial excitation on a proton given by
the diagrams in fig. \ref{vv}a to that for the ground state shown in fig.
\ref{vv}b
reads,

\beq
R_{V'/V}(Q^2) = \frac
{\displaystyle r_V \left(\frac{\Gamma_{V'}^{l\bar
l}m_V}{\Gamma_{V}^{l\bar l}m_V'}\right)^{1\over 2}
\left(\frac{1+Q^2/m_V^2}{1+Q^2/m_{V'}^2} \right) +
\epsilon_V}
{\displaystyle 1 + \epsilon_V \left(\frac{\Gamma_{V'}^{l\bar
l}m_V}{\Gamma_{V}^{l\bar l}m_V'}\right)^{1\over 2}
\left(\frac{1+Q^2/m_V^2}{1+Q^2/m_{V'}^2}
\right)}\ ,
\label{10}
\eeq
where $\Gamma^{l\bar l}$ is the partial leptonic width of the vector meson.

 To calculate the parameters $r_V$ and $\epsilon_V$ one needs a dynamical
 model. For heavy quarkonia one can use the pQCD-based
approach, developed in \cite{kz},

 \beq f_{V_2V_1}=
\langle V_2|\sigma(r_T)|V_1\rangle
\equiv \int d^3\vec r\
 V_2^*(\vec r)V_1(\vec r)\ \sigma(r_T)
\label{11}
\eeq

 Here $V_{1,2}(\vec r)$ are the quark wave function of the initial and the
final vector
 mesons. The flavour-independent dipole cross section, $\sigma(r_T)$,
of a $q\bar q$ pair interaction with a nucleon depends only
on the transverse separation $r_T$ \cite{zkl}.  Due to
 color screening $\sigma(r_T)\propto r_T^2$ at small $r_T$, what is a fairly
good
 approximation for heavy quarkonia.  Using the nonrelativistic oscillatory
model
 for the wave functions we get \cite{kz,hk} $\epsilon_V = - \sqrt{2/3}$
 (conventionally $V(0)$ and $V'(0)$ have the same sign) and $r_V = 7/3$.  Using
 these parameters and the data \cite{pdt}
 for $\Gamma_{\Psi'}^{l\bar l}/\Gamma_{\Psi}^{l\bar l}$
 we get from eq.  (\ref{10}) $R^2(Q^2=0) \approx 0.25$, what is quite close to
the
 measured value $R^2\approx 0.15 \pm 0.05$ in the real photoproduction
\cite{cam}.

 It is interesting that VDM in the form presented in fig. \ref{vv} predicts
according to eq.(\ref{10}) for $R^2_{\Psi'/\Psi}(Q^2)$ a
$Q^2$-dependence, similar to what follows from the pQCD based
calculations \cite{kz},

 \beq R^2(Q^2)=\left |\frac{\langle V'| \sigma(r_T)| \gamma^* \rangle} {\langle
 V| \sigma(r_T)| \gamma^* \rangle}\right |^2\ ,
\label{12}
\eeq The light-cone wave function of the $q\bar q$ component of the photon
reads
 \cite{nz91},

\begin{figure}[tbh]
\includegraphics{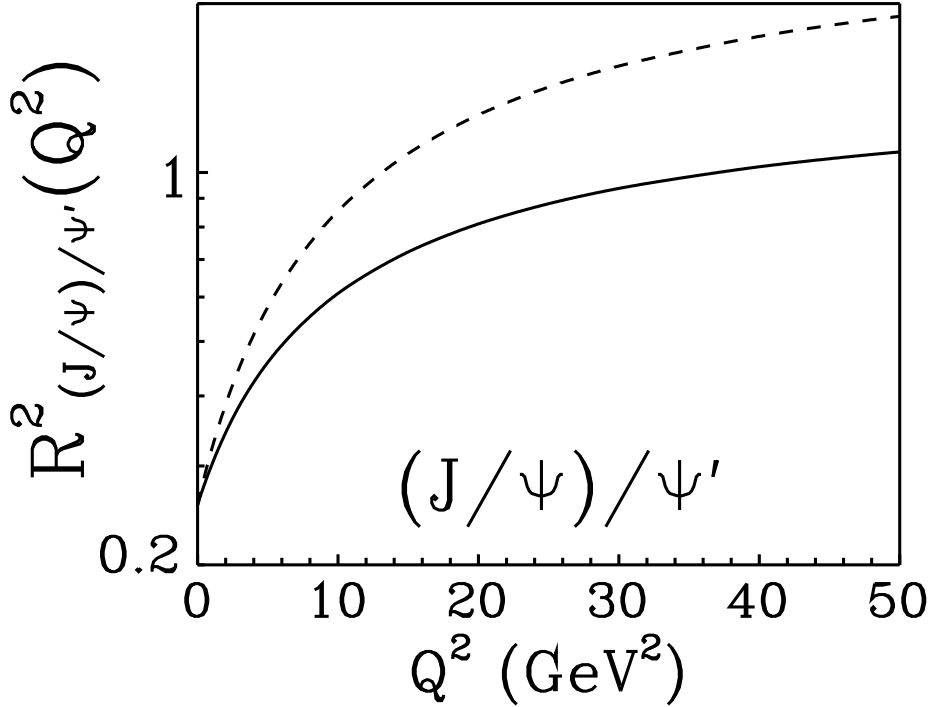}
\begin{center}
\vspace{9.5cm}
\parbox{13cm}
{\caption[Delta]
{$Q^2$-dependence of the ratio of the production rates of the
$J/\Psi$ to $\Psi'$ on a proton. The dashed and solid curves
correspond to eqs. (\ref{10}) and (\ref{12}) respectively}
\label{psi}}
\end{center}
\end{figure}

 \beq |\gamma^*\rangle \propto K_0[r_T\sqrt{m_q^2+Q^2/4}]\ ,
\label{13}
\eeq where $K_0$ is the modified Bessel function.

Results of calculation of $R_{\Psi'/\Psi}^2(Q^2)$ with eqs. (\ref{10}) and
 (\ref{12}) are presented in fig. \ref{psi}. The rising $Q^2$-dependence of
$R^2$ which
 follows from eq. (\ref{12}) is usually interpreted in terms of CT
 \cite{knnz}: the higher $Q^2$ is, the smaller is the mean transverse
 separation $\langle r_T^2\rangle$ in the produced $q\bar q$ wave packet, the
 less is the influence of the node in the wave function of the radial
 excitation which provides the suppression of the $V'$ photoproduction
 \cite{knnz}. We see that VDM results in a similar and even steeper growth of
 $R^2(Q^2)$, so an observation of such a behaviour cannot be interpreted as
 a confirmation of CT.

\medskip

 Situation with the light vector mesons ($\rho'(1450),\ \omega'(1420),\
\phi'(1680)$) is less certain. Firstly, the leptonic decay widths are poorly
known, or unknown at all. Secondly, the approximation $\sigma(r_T)\propto
r_T^2$ is not justified anymore, since large interquark distances are important
at low $Q^2$. Thirdly, the nonrelativistic oscillatory wave functions are
too rough approximation for light mesons.

Our results for the light vector mesons are summarized in Table 1.

\vspace{1cm}

\begin{tabular}{|c|c|l|l|l|l|c|}
\hline \multicolumn{6}{|c|}{\sc Table 1.
Ratio of the $V'/V$ production amplitudes}\\
 \hline Wave function & V-meson & r & $\epsilon$ &
 $\Gamma_{V'}^{l\bar l}/\Gamma_{V}^{l\bar l}$ & $R^2(V'/V)$\\
 \hline Nonrelativistic & $\rho$ & 1.5 & $-0.5$  &0.42 &
 0.074\\
\cline{2-6} Quark Model
& $\phi$ &1.6 &-0.56 &0.55 &0.290\\
 \hline
 Relativized & $\rho$ &1.25  & -0.14 &0.41 &
 0.22\\
 \cline{2-6} Quark Model
& $\phi$ &1.5 &-0.26 &0.38 &0.28\\
\hline Experimental & $\rho$ & &  &0.27-0.36 &
\\
\cline{2-6}
Data& $\phi$ & & &0.35 &\\
\hline
\end{tabular}

\vspace{1cm}

 In this case the large separations $r_T$ are important and we modify
$\sigma(r_T)$  at large distances to incorporate confinement as it is
suggested in \cite{nz91}.
 Using this dipole cross section and relativized wave functions of $\rho$- and
$\rho'$-mesons borrowed from ref.  \cite{nemchik} we calculated the parameters
$\epsilon_{\rho} = - 0.14$ and $r_{\rho} = 1.25$.

 Ratio of the leptonic widths is estimated in the relativistic quark model in
ref. \cite{gi} at $\Gamma^{l\bar l}_{\rho'}/\Gamma^{l\bar l}_{\rho} =
0.064$. Using these values of parameters we predict
$R^2_{\rho'/\rho}(Q^2=0)\approx 0.005$, what is at least an order
of magnitude lower than the
measured value \cite{solgado}.

 One can try to extract $\Gamma^{l\bar l}_{\rho'}$ using available data
\cite{pdt} for different decay channels of $\rho'$.  Combining information on
$\eta\rho$ and $\omega\pi$ channels we get $\Gamma^{l\bar l}_{\rho'}\approx
1.8\ KeV$.  The data \cite{pdt} for $\pi\pi$ and
$\omega\pi$ decay channels lead to $\Gamma^{l\bar l}_{\rho'}\approx 2.4\ KeV$.
As a rough estimate we fix the leptonic partial width at $2\
KeV$\footnote{similar value of $\Gamma^{l\bar l}_{\rho'}$ was found from
the analyses \cite{afs}}.  Then eq.  (\ref{10}) gives
$R^2_{\rho'/\rho}(Q^2=0)\approx 0.22$.  This value agrees with
available data for relative $\rho'/\rho$ photoproduction rates \cite{solgado},
provided that $2\pi^+2\pi^-$ branching is about $10\%$.

\begin{figure}[tbh]
\includegraphics{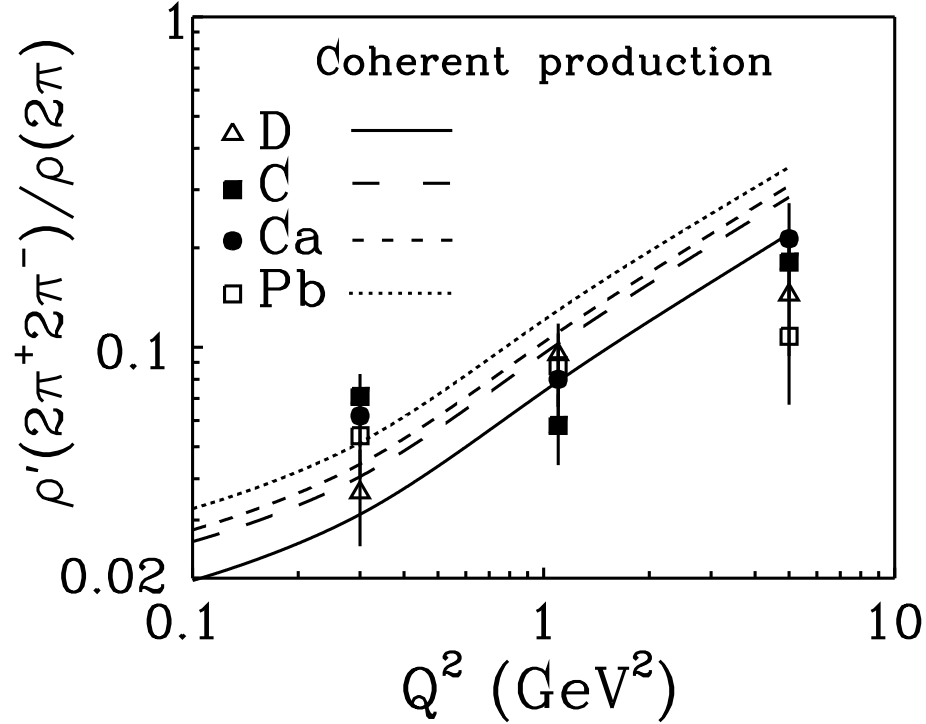}
\begin{center}
\vspace{9.5cm}
\parbox{13cm}
{\caption[Delta]
{$Q^2$-dependence of the ratio of coherent photoproduction
rates of $\rho'$ to $\rho$ on different nuclei. The curves
correspond to the two-channel VDM model incorporating
coherence length effects.
The data points are from the E665 experiment \cite{solgado}}
\label{solgado}}
\end{center}
\end{figure}

 Note, we predict the positive sign  for $R_{\rho'/\rho}(Q^2=0)$
what is different from pQCD estimation in refs.  \cite{knnz,nnz} which use
formula (\ref{12}).  In the latter case the negative sign of $R$ results from
"overcompensation" of the large-$r_T$ part of the $\rho'$ wave function
compared to the short distance part.  As a consequence, a change of sign of
$R(Q^2)$, i.e.  a sharp minimum in $Q^2$-dependence of $R^2$ at $Q^2\approx
0.3\ GeV^2$ was predicted in \cite{nnz}.
Such an approach is based on
the pQCD evaluation of the photon wave function. We work here in the
hadronic basis and use experimental data for most important
low-mass states.As is different from \cite{nnz} we
expect a smooth monotonous increase of
$R^2_{\rho'/\rho}$ as a function of $Q^2$ as is shown in fig. \ref{solgado}
 together with the
preliminary data from the E665 experiment
\cite{solgado}.
Again, the predicted $Q^2$-dependence of $R^2$ originates from
VDM and should not be misinterpreted as an evidence of CT.
Important is only the negative sign of $f_{VV'}$ which is
associated with any reasonable model, which  predicts
a rising $r_T$-dependence of $\sigma(r_T)$.

We use experimental information on
$\gamma \rightarrow \rho$ and $\gamma \rightarrow \rho'$ couplings, which seems
to
to provide more reliable predictions than that in \cite{knnz,nnz} where the
wave function of the $q\bar q$ fluctuation of the photon was estimated in the
nonrelativistic quark model neglecting the interquark interaction potential.

\medskip

 Using the same procedure we estimate the diffractive amplitudes for $\phi$
 and $\phi'$ interaction, $\epsilon_{\phi} = - 0.26,\ r_{\phi} = 1.5$.

\section{Nuclear effects in electroproduction of the
radial excitations}

 As soon as we included in the consideration production of the radial
excitations,
we are enforced to consider the production and the propagation through the
nucleus of a
wave packet which is a linear combination of $V$ and $V'$.
Corresponding approach was developed and used in refs.
\cite{kl,k-rev,jk,fgms,hk}.  The evolution equation for the wave packet
$|\Psi(z)\rangle =
{V\choose V'}$ propagating through the nuclear matter has a form of the
Schr\"odinger equation,

 \beq
i\frac{d}{dz}|\Psi(z,\vec b)\rangle= \widehat U(z,\vec b)|\Psi(z,\vec
 b)\rangle\ ,
\label{14}
\eeq

The evolution operator can be represented in the form $\widehat
 U=\hat q -i\hat f\ \rho(b,z)$, where

 \beq \widehat q=\left(\begin{array}{cc}q_L&0\\0&q_L'\end{array} \right)
\label{15}
\eeq

 \beq \widehat f=\frac{\stot}{2} \left(\begin{array}{cc}1&\epsilon\\\epsilon&r
\end{array}\right)
\label{16}
\eeq
All quantities here were defined above. except $q'_L = (Q^2+m_V'^2)/2\nu$.

It is interesting that the parameters $r$, $\epsilon$ and $R^2$ which
we found in the previous section for real photoproduction of charmonia, are
just
those magic values \cite{hk}, which correspond to the production in the initial
state
a combination of $J/\Psi$ and $\Psi'$, which is the eigen state of interaction.
Such a wave packet propagates through the nuclear matter with constant
relative content of $J/\Psi$ and $\Psi'$. This means, that nuclear
suppression is the same for $\Psi'$ and $J/\Psi$. This surprising effect was
observed
in hadroproduction of charmonia \cite{e772} and explained in \cite{hk}.
Now we expect the same for the real photoproduction.
A stronger relative enhancement of the  $\Psi'$ photoproduction on nuclei was
predicted
in \cite{kz}. That approach has an advantage of using the path-integral
technique, what is equivalent to the general multichannel consideration.
However it relays upon the projection to the quark wave function of
the photon in the form of eq. (\ref{13}). We use the experimental data
for $\Gamma^{l\bar l}_{\Psi'}/\Gamma^{l\bar l}_{J/\Psi}$ at this point.
Note, whatever prediction is correct, the principal is the coupled-channel
approach for propagation of the charmonium in the medium, first developed in
\cite{kz}.

\begin{figure}[tbh]
\includegraphics{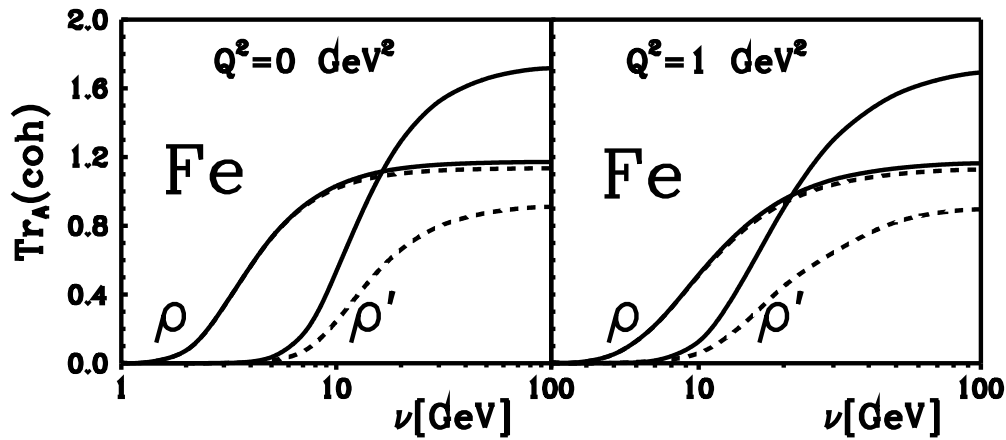}
\begin{center}
\vspace{6cm}
\parbox{13cm}
{\caption[Delta]
{Comparison of the Glauber model prediction (dashed curves) with the
prediction of the two-channel approach (solid curves) for
the nuclear transparency in the coherent electroproduction of the $\rho$-
and $\rho'$-mesons on iron as function of energy}
\label{rho}}
\end{center}
\end{figure}

\begin{figure}[tbh]
\includegraphics{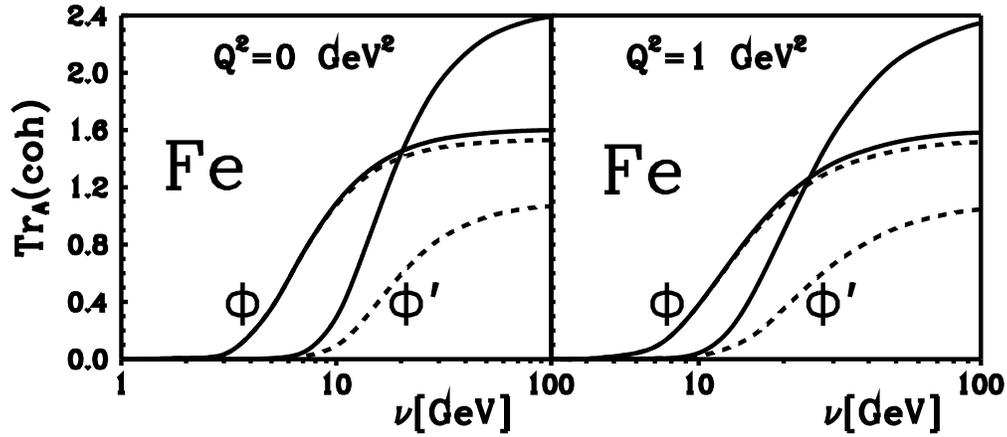}
\begin{center}
\vspace{6cm}
\parbox{13cm}
{\caption[Delta]
{The same as in fig. \ref{rho} for the electroproduction of
$\phi$- and $\phi'$-mesons}
\label{phi}}
\end{center}
\end{figure}

We have solved evolution equation (\ref{14}) numerically
for the realistic nuclear density distribution \cite{jager}. We used
the experimendal data for $\Gamma^{l\bar l}_{V'}/\Gamma^{l\bar l}_{V}$ and
the parameters $r$ and $\epsilon$ corresponding to the relativized
quark model in Table 1.  The results
for coherent production of $\rho,\ \rho'$ and $\phi,\phi'$
are shown by the solid curves in figs. \ref{rho} and  \ref{phi} respectively.
Predictions of the eikonal Glauber approximation are plotted by the dashed
curves
for comparison.

We see that the addition of the off-diagonal diffractive
amplitudes $V \rightleftharpoons V'$ to the eikonal
approximation results in a miserable correction to the
nuclear transparency for
the ground states. However, it leads to a dramatic
enhancement of nuclear transparency for the radial excitations.
This is easily interpreted as a result of the weakness
of the relative production rate of the radial excitations on a proton.
In this case the off-diagonal amplitude $\gamma \rightarrow V' \rightarrow
V$ turns out to be suppressed twice compared with
 $\gamma \rightarrow V \rightarrow V$, so the production of the ground states
gains
a tiny distortion. On the other hand, for production
of the radial excitation $V'$ the main correction comes from
the off-diagonal amplitude  $\gamma \rightarrow V \rightarrow V'$, which is
enhanced
compared to the diagonal one $\gamma \rightarrow V' \rightarrow V'$. This is
the main source of the abnormal nuclear enhancement of
the photoproduction of the radial excitations of the vector mesons,
discovered in \cite{kz}.
Note that this effect is usually considered
as a manifestation of CT effects, however, we see that it may have
quite a simple origin.

We have restricted our consideration in this section with a simplest case of
coherent productions. Evolution equation can be used for incoherent
production at the same footing, and its generalization to a multichannel
case is not difficult. Those results will be published elsewhere.

\section{Conclusions}

In present paper we consider the typical reactions
with electron beams, which are under an intensive discussion
recent years
as an effective way to search for CT.
However, we try here to be maximally free of
color dynamics and CT effects and we do that
{\it on purpose}. We either stay with the standard eikonal
Glauber model incorporated with the coherence length effects, or
we go beyond this approximation, but include only those
corrections which do not contain an explicit color dynamics.
This is to provide a better {\it baseline} for searching for CT effects.
We came to conclusion that at moderate electron energies the
standard mechanisms lead sometimes to the effects which are
usually expected
to be a signature of CT, or at least of its onset.

\medskip

If the authors of ref. \cite{murthy} had been asked $20$ years
ago to repeat their calculations, properly modified to predict $Q^2$-dependence
of
nuclear transparency in $(e,e'p)$ reaction, they would have provided
the prediction similar to what is depicted in fig. \ref{fig3}.
It is quite probable that such a behaviour will be observed soon in
future experiments at CEBAF, HERMES, ELFE. However, one should be cautious
interpreting it as a signal of CT.

Existing proposals \cite{cebaf,hermes} to study CT effects in
electroproduction of vector mesons at CEBAF or HERMES
rely upon the predictions \cite{knnz} which were done for
asymptotic-energies. However, the energy range of interest is
most sensitive to another effect associated with the energy- and
$Q^2$-variation of the coherence length. This results
in dramatic changes of the nuclear transparency even within
the standard Glauber approximation.
The formula for incoherent virtual photoproduction of vector
mesons, which is valid through the whole energy range, was unknown before.
We predict a variation of nuclear transparency up to a few hundred percents
for heavy nuclei, as a function of the energy or virtuality of the photon.
 These effects easily mock CT in the incoherent
productions, and anyway should be taken into account to
predict the baseline for CT.

In the case of virtual photoproduction of the radial excitations
we use the two-channel VDM approximation which provides a $Q^2$-dependence
of relative yields of $V'$ to $V$ and the nuclear antishadowing
similar to what is supposed to result from the color dynamics of
interaction, namely, from decreasing $Q^2$-dependence of the
transverse separation of the photoproduced $q\bar q$ wave packet.

\medskip

 {\bf Acknowledgement:} Most of presented results were obtained
in collaboration with J.~H\"ufner, whom we are grateful for
very helpful inspiring discussions. We are also thankful to H.~Pirner for
many useful and stimulating discussions.

\end{document}